\documentclass{aa}
\usepackage{times}
\usepackage{graphicx}
\usepackage{natbib}

\bibpunct{(}{)}{;}{a}{}{,}

\begin{document}

\def\aa{{A\&A}}
\def\aas{{ A\&AS}}
\def\aj{{AJ}}
\def\al{$\alpha$}
\def\bet{$\beta$}
\def\amin{$^\prime$}
\def\annrev{{ARA\&A}}
\def\apj{{ApJ}}
\def\apjs{{ApJS}}
\def\asec{$^{\prime\prime}$}
\def\baas{{BAAS}}
\def\cc{cm$^{-3}$}
\def\deg{$^{\circ}$}
\def\ddeg{{\rlap.}$^{\circ}$}
\def\dsec{{\rlap.}$^{\prime\prime}$}
\def\cc{cm$^{-3}$}
\def\e#1{$\times$10$^{#1}$}
\def\etal{{et al. }}
\def\flamb{erg s$^{-1}$ cm$^{-2}$ \AA$^{-1}$}
\def\flux{erg s$^{-1}$ cm$^{-2}$}
\def\fnu{erg s$^{-1}$ cm$^{-2}$ Hz$^{-1}$}
\def\hal{H$\alpha$}
\def\hst{{\it HST}}
\def\kms{km s$^{-1}$}
\def\lamb{$\lambda$}
\def\lax{{$\mathrel{\hbox{\rlap{\hbox{\lower4pt\hbox{$\sim$}}}\hbox{$<$}}}$}}
\def\gax{{$\mathrel{\hbox{\rlap{\hbox{\lower4pt\hbox{$\sim$}}}\hbox{$>$}}}$}}
\def\simlt{\lower.5ex\hbox{$\; \buildrel < \over \sim \;$}}
\def\simgt{\lower.5ex\hbox{$\; \buildrel > \over \sim \;$}}
\def\lum{erg s$^{-1}$}
\def\mbh{{$M_{\rm BH}$}}
\def\micron{{$\mu$m}}
\def\mnras{{MNRAS}}
\def\nat{{Nature}}
\def\pasp{{PASP}}
\def\perang{\AA$^{-1}$}
\def\percm2{cm$^{-2}$}
\def\peryr{yr$^{-1}$}
\def\pp{\parshape 2 0truein 6.1truein .3truein 5.5truein}
\def\reference{\noindent\pp}
\def\refindent{\par\noindent\parskip=2pt\hangindent=3pc\hangafter=1 }
\def\solum{$L_\odot$}
\def\solmass{$M_\odot$}
\def\oii{[\ion{O}{2}]}
\def\heii{\ion{He}{2}}
\def\hi{\ion{H}{1}}
\def\hii{\ion{H}{2}}
\def\oiii{[\ion{O}{3}]}
\def\ni{[\ion{N}{1}]}
\def\oi{[\ion{O}{1}]}
\def\nii{[\ion{N}{2}]}
\def\hei{\ion{He}{1}}
\def\sii{[\ion{S}{2}]}
\def\siii{[\ion{S}{3}]}
\def\lhal{$L_{{\rm H}\alpha}$}
\def\lbol{$L_{{\rm bol}}$}
\def\ledd{$L_{{\rm Edd}}$}

\def\bhm{M_{\rm BH}}
\def\cs{c_{\rm s}}
\def\dotm{\dot{m}}
\def\dotM{\dot{M}}
\def\dotMedd{\dot{M}_{\rm Edd}}
\def\dv{\Delta v}
\def\ledd{L_{\rm Edd}}
\def\lline{L_{\rm lines}}
\def\mbh{m_{\rm bh}}
\def\ooo{[O\,{\sc iii}]}

\def\ha{H$\alpha$}
\def\hb{H$\beta$}
\def\hg{H$\gamma$}
\def\civ{{\sc C iv}}
\def\mgii{Mg {\sc ii}}
\def\lya{Ly$\alpha$}

\title{The Central Engines of Radio-Loud Quasars}
\author{Jian-Min Wang\inst{1,2,3}
   \and Luis C. Ho\inst{4}
   \and R\"udiger Staubert\inst{1}}

\authorrunning{J.-M. Wang, L.C. Ho and R. Staubert}
\offprints{J.-M. Wang\\ \email{wang@astro.uni-tuebingen.de}}

\institute{Institut f\"ur Astronomie und Astrophysik, Abt. Astronomie,
Universit\"at T\"ubingen, Sand 1, D-72076 T\"ubingen, Germany,
\and Laboratory for High Energy Astrophysics, Institute of High
Energy Physics, CAS, Beijing 100039, P. R. China.
\and Alexander von Humboldt Fellow.
\and The Observatories of the Carnegie Institution of Washington,
813 Santa Barbara Street, Pasadena, CA 91101-1292; lho@ociw.edu.}

\date{Received $<$date$>$ / Accepted $<$date$>$}

\abstract{
We have assembled a sample of 37 radio-loud quasars that have been
imaged with the {\it Hubble Space Telescope}\ in order to investigate
their black hole masses, accretion rates, and the structure of their
accretion disks.  The black hole masses were estimated from the
luminosities of the host galaxies, and the accretion powers were extrapolated
from the emission-line luminosities.  The majority of the quasars have masses
in the range $M_{\rm BH} \approx 10^8-10^9$ \solmass.  Their accretion rates,
$\dot M \approx 0.01-1$ times the Eddington rate, suggest that
most of the objects possess standard optically thick, geometrically thin
accretion disks, in some cases perhaps accompanied by an optically thin
advection-dominated component.  The coexistence of strong radio emission
and a standard disk conflicts with recent models for jet formation.  We
discuss modifications of the standard model that can resolve this discrepancy.
We find there is a strong correlation between the accretion rate and
the extended radio luminosity. This lends support to the idea that the extended
radio emission is somehow linked to the accretion disk.  Lastly, we combine
the present sample of radio-loud quasars with the sample of BL Lac objects
studied by Wang, Staubert, \& Ho (2002) to reevaluate the unification picture
for radio-loud active galactic nuclei.  Consistent with
current ideas for the unification of radio-loud sources, we
find that flat-spectrum radio quasars and FR~II radio galaxies indeed
seem to belong to the same population, as do BL Lac objects and FR~I radio
galaxies on the opposite end of the luminosity spectrum.  However, some
members of the low frequency-peaked BL Lac objects may be more closely
associated with FR~II rather than FR~I radio galaxies.  We describe how the
various subclasses of radio-loud sources can be viewed as a continuous
sequence of varying accretion rate.%
}

\maketitle

\section{Introduction}

According to the radio-loudness criterion proposed by Kellermann et al. (1989),
quasars whose ratio of radio (5 GHz) to optical ($B$ band) flux exceeds 10
comprise a special population of radio-loud objects (hereafter RLQs).  As
reviewed by Urry \& Padovani
(1995), the powerful radio emission is thought to originate from relativistic
jets expelled by the accretion disk around a central supermassive black hole
(BH).  One of the unsolved problems in astrophysics is the exact physical
relation between the jet and the accretion disk.  Two observational approaches
have been taken to investigate this relation in RLQs and active galactic
nuclei (AGNs). On large (kpc to Mpc) scales the relation between the jet 
radio/kinetic luminosity and the narrow-line luminosity has been examined by 
many authors (e.g.,  Baum \& Heckman 1989a, 1989b;
Rawlings \& Saunders 1991). It has been found that the narrow-line luminosity
is correlated with the kinetic luminosity from the radio lobes.  On
smaller scales, Celotti \& Fabian (1993) and Celotti, Padovani, \& Ghisellini
(1997) used radio data from very long-baseline interferometry to estimate
the kinetic luminosities in a large sample of radio-loud objects. Both
attempts reached the same conclusion:  the relativistic jet is linked
with the accretion disk. A more direct connection between the jet and the
accretion disk has be found in the RLQ 3C 120 (Marscher et al.
2002). However, how a relativistic jet actually forms from the accretion disk
remains an outstanding puzzle (Meier 2001; Blandford 2002).

\newdimen\digitwidth    
\setbox1=\hbox{0}       
\digitwidth=\wd1        
\catcode`"=\active      
\def"{\kern\digitwidth}
\begin{table*}
\footnotesize
\centerline{{\bf Table 1.} The Sample of Radio-Loud Quasars}
\scriptsize
\begin{center}
\begin{tabular}{lllccllccc}\\ \hline \hline
No.&Object&Name&Type&$z$&$M_R$(host)&$M_R$(QSO)&Reference&$L_{\rm 5 GHz}^{\rm ext}$&~~Reference\\
(1)&~~~(2) &~~(3)&(4) &(5)&~~~~~~~~(6)      &~~~~~~~~(7) &(8)  &(9) &~~~~~(10)\\ \hline
1 &0133$+$207"&3C 47    &S & 0.425&$-$23.57 &$-$24.27 &2 &26.71&6 \\
2 &0137$+$012"&PHL 1093 &S & 0.258&$-$23.74 &$-$23.67 &2 &27.07&7 \\
3 &0137$+$3309&3C 48    &S & 0.367&$-$25.03 &$-$25.92 &2 &27.52&6 \\
4 &0202$-$765"&PKS      &S & 0.389&$-$23.32 &$-$25.18 &2 &...  &   \\
5 &0312$-$770"&PKS      &F & 0.223&$-$23.98 &$-$24.60 &2 &...  &   \\
6 &0340$+$048"&3C 93    &S & 0.357&$-$23.93 &$-$23.71 &2 &25.56&8 \\
7 &0405$-$123"&OF$-$109 &F & 0.573&$-$23.97 &$-$27.87 &1 &27.16&6 \\
8 &0454$-$22""&PKS      &S & 0.533&$-$21.39 &$-$26.79 &1 &27.57&9 \\
9 &0736$+$01""&OI+061   &F & 0.191&$-$23.78 &$-$24.23 &2 &24.32&6 \\
10&0837$-$120"&3C 206   &S & 0.198&$-$23.29 &$-$24.23 &2 &26.03&6 \\
11&0903$+$169"&3C 215   &S & 0.412&$-$23.28 &$-$24.28 &2 &26.51&6 \\
12&1004$+$130"&4C 13.41 &S & 0.240&$-$24.26 &$-$25.82 &2 &26.15&6 \\
13&1020$-$103"&OL$-$133 &S & 0.197&$-$23.36 &$-$23.47 &3 &24.83&13 \\
14&1138$+$0003&LBQS     &? & 0.500&$-$23.97 &$-$24.35 &4 &...  &   \\
15&1217$+$023"&ON+029   &F & 0.239&$-$23.71 &$-$24.40 &3 &25.45&6 \\
16&1218$+$1734&LBQS     &~~S$^a$ & 0.445&$-$23.46 &$-$23.74 &2 &25.93&14  \\
17&1222$+$1235&LBQS     &~~S$^b$ & 0.412&$-$24.02 &$-$24.24 &2 &25.96&14  \\
18&1226$+$023"&3C 273   &F & 0.158&$-$24.44 &$-$27.39 &2 &27.04&6 \\
19&1230$-$0015&LBQS     &S & 0.470&$-$23.14 &$-$24.64 &4 &...  &   \\
20&1250$+$568"&3C 277.1 &S & 0.321&$-$22.96 &$-$23.29 &2 &26.73&6 \\
21&1302$-$102"&OP$-$106 &S & 0.286&$-$24.34 &$-$26.13 &2 &25.70&6 \\
22&1309$+$355"&Ton 1565 &F & 0.184&$-$23.82 &$-$24.73 &2 &24.14&6 \\
23&1425$+$267"&Ton 202  &S & 0.366&$-$23.65 &$-$25.69 &2 &25.78&6 \\
24&1512$+$37""&4C 37.43 &S & 0.371&$-$23.86 &$-$25.58 &2 &26.28&6 \\
25&1545$+$210"&3C 323.1 &S & 0.266&$-$23.53 &$-$25.14 &2 &26.42&6 \\
26&1548$+$114A&4C+11.50 &F & 0.436&$-$22.40 &$-$23.86 &2 &25.06&10 \\
27&1641$+$399"&3C 345   &F & 0.593&$-$23.75 &$-$26.35 &1 &26.48&11 \\
28&1704$+$608"&3C 351   &S & 0.371&$-$24.79 &$-$26.16 &2 &26.76&6 \\
29&2128$-$123"&OX$-$148 &F & 0.501&$-$22.82 &$-$27.11 &5 &...  &   \\
30&2135$-$147"&PHL 1657 &S & 0.200&$-$23.43 &$-$24.16 &2 &26.34&6 \\
31&2141$+$175"&OX 169   &F & 0.213&$-$23.38 &$-$24.79 &2 &26.33&6 \\
32&2201$+$315"&4C 31.63 &F & 0.295&$-$24.70 &$-$25.98 &2 &26.54&6 \\
33&2247$+$140"&4C 14.82 &S & 0.237&$-$23.52 &$-$24.10 &2 &...  &   \\
34&2348$+$0210&LBQS     &? & 0.504&$-$23.75 &$-$25.13 &4 &...  &   \\
35&2349$-$014"&PB 5564  &S & 0.173&$-$24.27 &$-$24.02 &2 &...  &   \\
36&2351$-$0036&LBQS     &F & 0.460&$-$23.02 &$-$24.18 &4 &...  &   \\
37&2355$-$082"&PHL 6113 &S & 0.210&$-$23.62 &$-$23.06 &3 &28.01&12 \\ \hline
\end{tabular}
\vskip 2pt
\parbox{5.0in}
{\small\baselineskip 9pt
\scriptsize
\indent
{\sc Notes---} Col. (1) number; col. (2): object; col. (3): common name; col.
(4): type (based on the spectral index between 1.4 and 5 GHz;
col. (5): redshift; col. (6) $R$-band absolute magnitude of the host galaxy;
col. (7): $R$-band absolute magnitude of the QSO; col. (8): references
for $M_R$; col. (9): extended radio power (in units of W~Hz$^{-1}$) at 5 GHz;
col. (10): references for $L_{\rm 5 GHz}^{\rm ext}$. \\
{\sc References---}
(1) Boyce, Disney, \& Bleaken 1999;
(2) Hamilton, Casertano, \& Turnshek 2002;
(3) Dunlop et al. 2003;
(4) Hooper, Impey, \& Foltz 1997;
(5) Boyce et al. 1998;
(6) Xu, Livio, \& Baum 1999;
(7) Rector, Stocke, \& Ellingson 1995;
(8) Bogers et al. 1994;
(9) Aldcroft, Elvis, \& Bechtold 1993;
(10) Hutchings et al. 1996;
(11) White 1992;
(12) Lister, Gower, \& Hutchings 1994;
(13) Brotherton 1996.
(14) Hooper, Impey, Foltz, \& Hewett 1996\\
$^a$ based on the index between 0.365 and 4.85GHz (Gregory \& Condon 1991 and
       Douglas et al. 1996)\\
$^b$ based on the index between 4.85 and 8.4 GHz (Gregory \& Condon 1991 and
       Ref. 14).

}
\end{center}
\vglue-0.5cm
\end{table*}
\normalsize

The situation in Galactic BH candidates suggests that the formation of the jet
is somehow linked with the accretion rate. The radio emission from these
objects becomes quenched when the system is in a soft (high accretion rate)
state, as seen in GX~339$-$4 (Fender et al. 1999), GRS~1915+105
(Harmon et al. 1997; Fender et al. 1999), GRO~J16655$-$40 (Harmon et al.
1995), Cynus~X-3 (McCollough et al. 1999), and Cygnus~X-1 (Brocksopp et al.
1999).  These objects may be instructive for the case of AGNs, although we
should be cautious that some of their properties do not scale linearly with
BH mass.  The relation between the ``states'' of the accretion disk and jet
production in supermassive BH accretion systems is very poorly understood.

There are two impediments in solving this problem. First, the mass of the
central BH is highly uncertain. BH masses derived from fitting the ``big blue
bump'' (Shields 1978; Malkan \& Sargent 1982), for example, are coupled to the
accretion rate, which is not known independently.  Second, the observed
continuum of RLQs, even at optical and ultraviolet wavelengths, may be
significantly contaminated by nonthermal emission from the relativistic
jet. With the exception of objects with prominent big blue bumps, it is
nontrivial to estimate the thermal component of the continuum.  Thus, it is
difficult to reliably deduce the state of the accretion disk, which can
be described conveniently by the dimensionless accretion rate
$\dotm = \dotM/\dot{M}_{\rm Edd}$, where $\dotM$ is the accretion rate,
$\dot M_{\rm Edd} = 2.2\times 10^{-8} \left(\eta/0.1\right)^{-1}
\left(M_{\rm BH}/M_{\odot}\right)$ \solmass\ \peryr\ is the Eddington accretion
rate defined by $\dot{M}_{\rm Edd}\,=\,L_{\rm Edd}/\eta c^2$,
$L_{\rm Edd}\,=\,1.26 \times 10^{38} \left(M_{\rm BH}/M_{\odot}\right)$ \lum\
is the Eddington luminosity, and $\eta$ is the radiative efficiency (Frank,
King, \& Raine 1992).

The situation has improved dramatically with the availability of {\it Hubble
Space Telescope (HST)}\ observations of quasar host galaxies.  Relatively
reliable luminosities for the host galaxies are now available.   From
studies of nearby inactive galaxies, it has been established that the mass of
the BH scales roughly linearly with the luminosity (or mass) of the bulge
component of the host galaxy (Kormendy \& Richstone 1995; Magorrian et al.
1998; Ho 1999; Kormendy \& Gebhardt 2001).  Thus, we can use the luminosities
of the quasar host galaxies, which are largely dominated by the bulge
component, to estimate BH masses for quasars.

\begin{figure*}[t]
\centerline{\includegraphics[angle=-90,width=16cm]{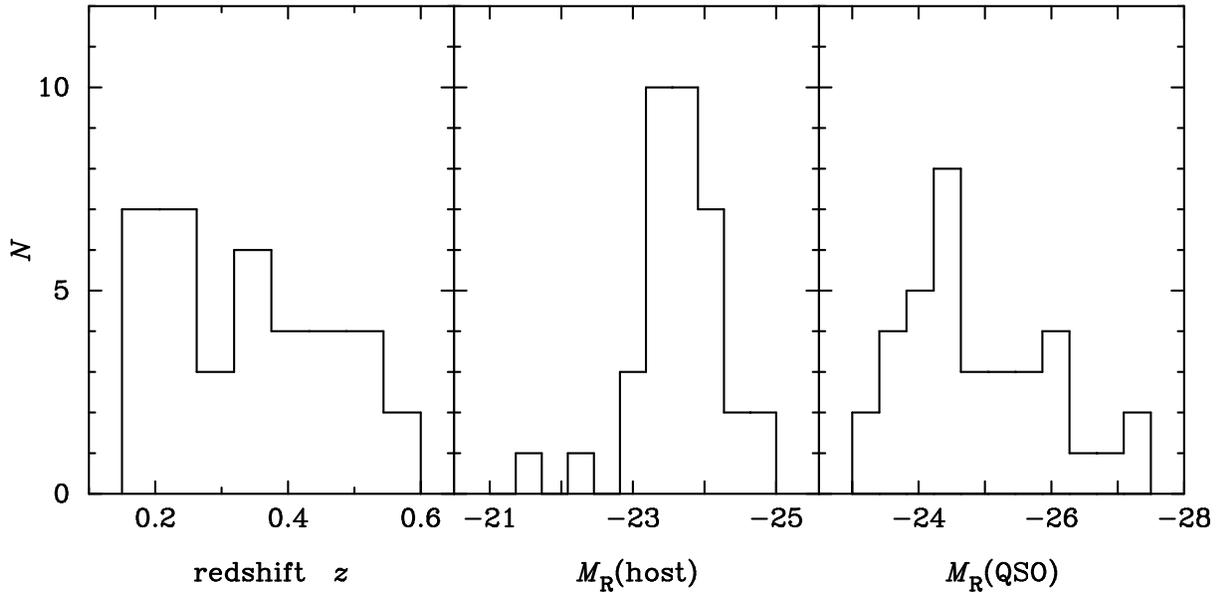}}
\caption{\footnotesize
Distribution of redshift, $M_R$(host), and $M_R$(QSO) for the
present sample.}
\label{fig1}
\centerline{}
\vglue -0.6cm
\end{figure*}

Accretion disks have complicate structures for different $\dot{m}$ and the
viscosity $\alpha$ (Chen et al. 1995). For a disk with a low accretion rate,
three possible solutions have been suggested: optically thin advection-dominated
accretion flows (ADAF; Narayan \& Yi 1994), advection-dominated inflow-outflows
(ADIOs; Blandford \& Begelman 1999) and convection-dominated accretion flows
(CDAF; Narayan, Igumenshchev \& Abramowicz 2000; Quataert \& Gruzinov 2000).
Whether low $\dot{m}$ necessarily leads to low $\eta$, however, remains
controversial.
When $\alpha^2<\dot{m}<1$, the disk has a standard optically thick,
geometrically thin structure (Shakura \& Sunyaev 1973, hereafter SS),
which has been extensively studied. When
$\dot{m}>1$, an optically thick ADAF or super-Eddington accretion flow (SEA)
with slim geometry is formed (Begelman \& Meier 1982; Abramowicz et al. 1988;
Chen \& Taam 1993). Such a disk has two possible structures: an inhomogeneous
structure (split into rings) due to the photon-bubble instability (Begelman
2002), or a homogeneous disk with photon trapping (Wang \& Zhou 1999; Ohsuga
et al. 2002). Once the mass of the BH and the accretion luminosity are known,
we can constrain the state of the accretion disk, which provides possible
clues to jet formation.

Recently, a number of authors have suggested that flat-spectrum radio
quasars, based on their very high kinetic luminosities, are powered by
super-Eddington accretion flows (Ghisellini \& Celotti 2001; Cavaliere \&
D'Elia 2002; Maraschi \& Tavecchio 2003).  It would be highly desirable to
test this and other hypotheses by obtaining a reliable estimate of the
accretion rates of RLQs.  This is the purpose of this paper.

\section{The Sample}

The most reliable measurement of quasar host galaxy luminosities currently
come from {\it HST}\ observations.  Our analysis uses RLQs with available
{\it HST}\ imaging, based mainly on three samples (see Table 1): (1) Hamilton
et al. (2002), (2) McLure et al. (2001) and Dunlop et al. (2003), and
(3) Boyce et al. (1998, 1999). The $B$ and $V$ magnitudes are converted into
the $R$ band following Fukugita, Shimasaku, \& Ichikawa (1995).
We adopt the results from Hamilton et al. (2002)
if the objects have been measured several times in order to decrease the
systematic errors.  Unfortunately, systematic errors in host galaxy
measurements are difficult to quantify, but uncertainties on the order
of 0.1--0.3 mag are not unexpected.
Table 1 also lists the absolute magnitudes of the quasar nuclei,
based on spatial decomposition of the {\it HST}\ images.  Finally, we give
published values of the extended radio luminosity.
Following standard practice, the objects are divided into two
subclasses according to their radio spectral index $n$, defined by
$F_{\nu}\propto \nu^n$: flat-spectrum radio quasars (FSRQs) are those with
$n\geq -0.5$, whereas steep-spectrum radio quasars (SSRQs) have $n<-0.5$,
where $n$ is generally calculated between 1.4 and 5 GHz.
There are two objects without spectral index, but 
they do not change our conclusions (see Table 2 and 3).

The redshift range of the sample is $0.16\le z\le 0.57$, and the host galaxy
absolute magnitudes span $-21.4\ge M_R\ge -25.0$ (Fig. 1).  For our present
application, we use the empirical BH mass-bulge luminosity relation of
McLure \& Dunlop (2002):
\begin{equation}
\log\left(M_{\rm BH}/M_{\odot}\right)=-0.5M_R-2.96.
\end{equation}
For ease of comparison with literature results, distance-dependent quantities
assume a Hubble constant of $H_0$ = 50 \kms\ Mpc$^{-1}$ and a deceleration
parameter of $q_0=0.5$.

\section{States of the Accretion Disk}

The observed optical--UV continuum of RLQs may be
contaminated by nonthermal emission from the relativistic jet, rendering it
difficult to estimate the thermal contribution to the continuum, and hence
to estimate accretion rates.  On the other hand, quasars, independent of their
radio strength, usually have strong, well-observed broad emission lines, which
leads to the possibility of using the broad-line region (BLR) emission-line
strength to indirectly estimate the ionizing luminosity.  Following Celotti
et al. (1997), this is the approach we will take here.  Celotti et al. (1997)
showed that the BLR luminosities in RLQs are roughly equal to the kinetic
luminosities of their jets. 
We assume that the BLR clouds are most likely photoionized by the thermal
emission from the accretion disk, rather than by nonthermal emission from the
jet. This assumption is supported by that
the beamed radiation from the jet has great difficulties to
ionize the clouds distributed within a large solid angle to produce a reasonable 
range of ionization parameter for emission lines.
There is also ample evidence that BLR emission lines are
photoionized by a thermal continuum (see, e.g., Netzer 1990). The underlying
assumption here should be 

\begin{table*}
\footnotesize
\centerline{{\bf Table 2.} Line Luminosities}
\scriptsize
\begin{center}
\begin{tabular}{lllcc}\\ \hline \hline
No.& Lines & Reference &log Flux (\flux) & log Luminosity (\lum)\\
~       &   ~   & ~    & &\\
(1) &~ (2)   &~(3)  &(4) & (5)\\ \hline
1 &\civ, \lya                 &15          &$-$11.74&  45.22\\
2 &\ha, \hb, \hg              &1           &$-$11.42&  45.09\\
3 &\hg, \civ, \mgii, \lya     &16          &$-$11.49&  45.34\\
4 &\hb,\hg, \mgii             &5           &$-$11.93&  44.95\\
5 &\civ, \lya                 &7,11        &$-$11.41&  44.96\\
6 &\ha                        &6           &$-$12.60&  44.21\\
7 &\ha, \hb, \civ, \mgii, \lya&7,9,10,11,12&$-$11.21&  46.03\\
8 &\hb, \civ, \lya            &9,11        &$-$11.64&  45.53\\
9 &\ha, \hb, \hg, \civ, \lya  &8,11        &$-$11.80&  44.43\\
10&\ha, \hb, \hg, \civ, \lya  &8,10,11     &$-$11.64&  44.63\\
11&\hg, \mgii                 &16          &$-$11.95&  44.98\\
12&\ha, \hb, \civ, \mgii, \lya&8,10,11     &$-$11.67&  44.77\\
13&\ha, \hb, \hg              &13          &$-$11.79&  44.47\\
15&\ha, \hb, \hg, \civ, \lya  &8,11,7      &$-$11.87&  44.57\\
18&\hb, \hg, \civ, \mgii, \lya&3,14        &$-$10.28&  45.79\\
20&\ha, \hb, \hg, \civ, \mgii &8,15        &$-$11.96&  44.75\\
21&\hb, \civ, \lya            &7,9,11      &$-$11.51&  45.09\\
22&\civ, \lya                 &11          &$-$11.50&  44.70\\
23&\ha, \hb, \civ, \mgii, \lya&10,11       &$-$11.66&  45.16\\
24&\ha, \hb, \civ, \mgii, \lya&7,9,10,11   &$-$11.59&  45.25\\
25&\hb, \civ, \mgii, \lya     &7,9,15      &$-$11.11&  45.42\\
26&\hb, \hg                   &2           &$-$12.34&  44.64\\
27&\hb, \hg, \civ, \lya       &8,11        &$-$11.84&  45.44\\
28&\ha, \hb, \civ, \mgii, \lya&7,9,10,11   &$-$11.63&  45.21\\
29&\hb, \hg, \civ, \mgii, \lya&2,7,11,12   &$-$12.04&  45.08\\
30&\hb, \civ, \lya            &7,9,11      &$-$11.40&  44.87\\
31&\hb, \civ, \lya            &7,9,11      &$-$11.69&  44.64\\
32&\ha, \hb, \hg, \civ, \lya  &7,8,9,10,11 &$-$11.31&  45.31\\
33&\hb                        &4           &$-$12.22&  44.21\\
35&\hb                        &4           &$-$11.25&  44.89\\ \hline
\end{tabular}
\vskip 2pt
\parbox{4.6in}
{\small\baselineskip 9pt
\scriptsize
\indent
{\sc Notes---}
Col (1) number; col (2) emission lines; col (3) references; col (4) the
derived total line fluxes; and col (5) the derived total line luminosity.\\
{\sc References---}
 (1) Baldwin 1975;
 (2) Bergeron \& Kunth 1984;
 (3) Corbin 1992;
 (4) Corbin 1997;
 (5) Danziger \& Goss 1983;
 (6) Eracleous \& Halpern 1994;
 (7) Gondhalekar 1990;
 (8) Jackson \& Browne 1991;
 (9) Marziani et al. 1996;
(10) Neugebauer et al. 1979;
(11) Osmer, Porter, \& Green 1994;
(12) Oke, Shields, \& Korycanski 1984;
(13) Scarpa \& Falomo 1997;
(14) Stockton \& MacKenty 1987;
(15) Wills et al. 1995;
(16) Wilkes et al. 1999.}

\end{center}
\vglue-0.5cm
\end{table*}
\normalsize

\noindent
quite safe in statistical test.

Table 2 lists emission-line fluxes we were able to assemble for our sample.
We choose prominent broad emission lines, such as H$\alpha$, H$\beta$,
H$\gamma$, \civ, \mgii~ and \lya, that account for a large fraction of the
total line luminosity.  The overall similarity of the emission-line spectra of
quasars suggests that the underlying physical conditions of their BLR are
quite similar (Netzer 1990; Francis et al.  1991; Boroson \& Green 1992; Zheng
et al. 1997). To first order, therefore, the composite spectrum of quasars
(Francis et al. 1991) may be used as a reasonable template to calculate the
total BLR luminosity (Celotti et al.  1997).  Then, from the relative
contribution of individual lines to the total BLR luminosity,
we obtain the total line luminosity $\lline$ (Table 2).

The fraction $\xi$ of the thermal disk emission reprocessed by the BLR clouds
is roughly equal to the cloud covering factor.  This assumption should be
statistically valid for large samples since in steady state the energy
absorbed by the BLR clouds is approximately that radiated as line emission.
Thus, the thermal emission from the accretion disk is
\begin{equation}
L_{\rm disk}=\xi^{-1}L_{\rm lines}.
\end{equation}
According to Netzer (1990), $\xi \approx 0.1$.

We recognize that, for any individual object, variability and internal
reddening may introduce uncertainties into $\lline$.  Source-to-source
variations in the covering factor or reprocessing efficiency may also add
additional scatter.
It is hoped, however, that the overall statistics of a large sample will not
be severely affected and that they are at least internally consistent.

Following Wang, Staubert, \& Ho (2002), we define the line accretion rate and
its dimensionless form as
\begin{equation}
\Lambda=\frac{\lline}{c^2};~~~~~\lambda=\frac{\lline}{L_{\rm Edd}}.
\end{equation}
The relation between $\lambda$ and the dimensionless accretion rate $\dot{m}$,
for an optically thin ADAF, is given by (Wang et al. 2002)
\begin{equation}
\dot{m}=2.17\times 10^{-2}\alpha_{0.3}\xi_{-1}^{-1/2}\lambda_{-4}^{1/2}.
\end{equation}
This expression assumes that the total luminosity from the ADAF is
$L_{\rm disk}\propto \alpha^{-2}M_{\rm BH} \dot{m}^2$ (Mahadevan 1997),
$\alpha_{0.3}=\alpha/0.3$, $\xi_{-1} = \xi/0.1$, and $\lambda_{-4} =
\lambda/10^{-4}$.  A necessary condition for the presence of an optically
thin ADAF is $\dot{m}\le \alpha^2$ (Narayan, Mahadevan, \& Quataert 1998).
Equation (4) can then be rewritten as
\begin{equation}
\lambda_{\rm 1}=1.72\times 10^{-3}\xi_{-1}\alpha_{0.3}^2.
\end{equation}
Optically thin ADAFs require $\lambda < \lambda_1$.

Accretion disks in the optically thick, geometrically thin SS regime
obey
\begin{equation}
\dot{m}=\frac{L_{\rm line}}{\xi L_{\rm Edd}}=10\xi_{-1}^{-1}\lambda,
\end{equation}
and the condition $1 > \dot{m} \ge \alpha^2$ gives
\begin{equation}
\lambda_2=9.0\times 10^{-3}\xi_{-1}\alpha_{0.3}^2,
\end{equation}
above which a standard disk can exist.  It is interesting to note that in
the transition region between $\lambda_1$ and $\lambda_2$ the accretion flow
may be in a hybrid state in which a standard disk coexists with an ADAF. The
possibility of hybrid states in AGN accretion disks has been discussed by,
among others, Quataert et al. (1999), R\'ozanska \& Czerny (2000), Ho et al.
(2000), and Ho (2002a).  A transition
from an SS disk to an ADAF is possible (Gu \& Lu 2000), perhaps via evaporation
(Liu et al. 1999).  The transition radius depends on the accretion rate,
the BH mass, and viscosity.  The disk structure, however, is complicated in
such a regime, largely due to uncertainties in the viscosity.

\begin{figure}
\centerline{\includegraphics[angle=-90,width=8.cm]{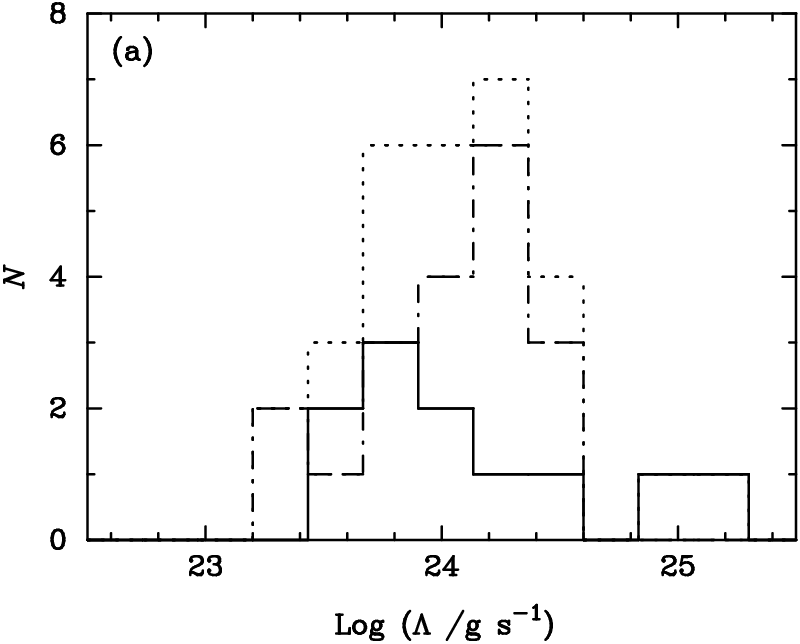}}
\vglue 0.5cm
\centerline{\includegraphics[angle=-90,width=8.cm]{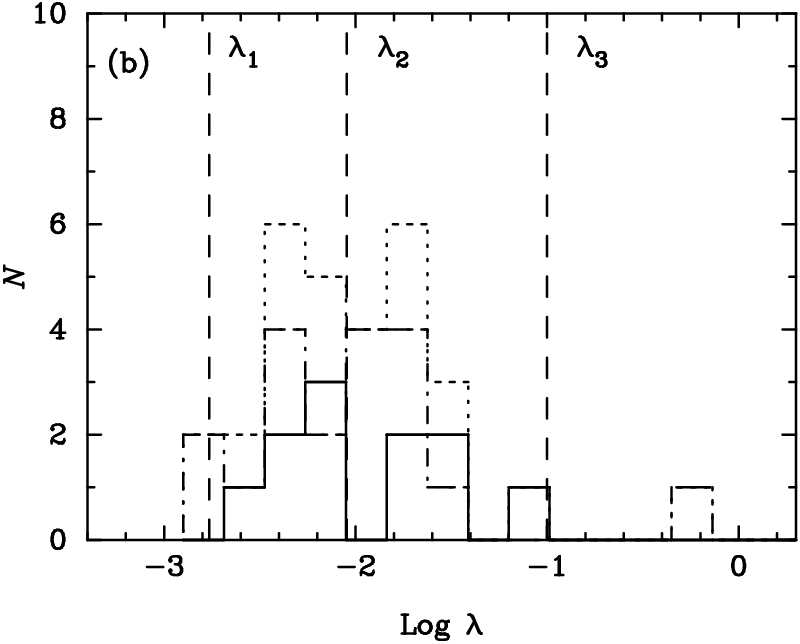}}
\caption{\footnotesize
The distribution of ({\it a}) $\Lambda$ and ({\it b}) $\lambda$.
The solid histograms plot the FSRQs, and the dotted-dashed histograms show
the SSRQs. The dotted histograms show the total distribution for all
RLQs (FSRQs and SSRQs).  The distribution of $\lambda$ is divided into four
regions, corresponding to different states of the accretion disks:
(1) $\lambda < \lambda_1$ (pure ADAF);
(2) $\lambda_1 \le \lambda \le \lambda_2$ (ADAF+SS);
(3) $\lambda_2 \le \lambda \le \lambda_3$ (SS);
(4) $\lambda \ge \lambda_3$ (SEA).
}
\label{fig2}
\end{figure}
When the accretion rate reaches $\dot{m}\ge 1$, we have
\begin{equation}
\lambda_3=0.1\xi_{-1},
\end{equation}
in which case we use
\begin{equation}
E_{\rm R}=\frac{L_{\rm line}}{\xi L_{\rm Edd}}
\end{equation}
\begin{center}
\footnotesize
\centerline{{\bf Table 3.} Results}
\scriptsize
\vglue -0.1cm
\begin{tabular}{lccll}\\ \hline \hline
Object& log $M_{\rm BH}$&log $\Lambda$& log $\lambda$& Disk\\
      & (\solmass) & (${\rm g~s^{-1}}$) & & \\ \hline
  1&  8.82&  24.27&-1.71&SS         \\
  2&  8.91&  24.14&-1.92&SS         \\
  3&  9.56&  24.39&-2.32&SS+ADAF    \\
  4&  8.70&  24.00&-1.85&SS         \\
  5&  9.03&  24.01&-2.17&SS+ADAF    \\
  6&  9.01&  23.26&-2.90&ADAF       \\
  7&  9.02&  25.08&-1.10&SS         \\
  8&  7.73&  24.58&-0.31&SEA        \\
  9&  8.93&  23.48&-2.60&SS+ADAF    \\
 10&  8.69&  23.68&-2.16&SS+ADAF    \\
 11&  8.68&  24.03&-1.80&SS         \\
 12&  9.17&  23.82&-2.50&SS+ADAF    \\
 13&  8.72&  23.52&-2.35&SS+ADAF    \\
 15&  8.89&  23.62&-2.43&SS+ADAF    \\
 18&  9.26&  24.84&-1.57&SS         \\
 20&  8.52&  23.80&-1.87&SS         \\
 21&  9.21&  24.14&-2.22&SS+ADAF    \\
 22&  8.95&  23.75&-2.35&SS+ADAF    \\
 23&  8.86&  24.21&-1.81&SS         \\
 24&  8.97&  24.30&-1.82&SS         \\
 25&  8.81&  24.47&-1.49&SS         \\
 26&  8.24&  23.69&-1.70&SS         \\
 27&  8.91&  24.49&-1.58&SS         \\
 28&  9.44&  24.26&-2.33&SS+ADAF    \\
 29&  8.45&  24.13&-1.47&SS         \\
 30&  8.76&  23.92&-1.99&SS         \\
 31&  8.73&  23.69&-2.19&SS+ADAF    \\
 32&  9.39&  24.36&-2.18&SS+ADAF    \\
 33&  8.80&  23.26&-2.69&SS+ADAF    \\
 35&  9.18&  23.94&-2.39&SS+ADAF    \\ \hline
\end{tabular}
\end{center}
\normalsize

\noindent
to describe the accretion state, since $\dot{m}>E_R$ in the super-Eddington
regime in which photon trapping efficiently lowers 
the radiated luminosity (Wang \& Zhou 1999; Ohsuga et al. 2002).

Our estimation of the parameter $\lambda$ are subject to at least
three sources of uncertainty.  First, the total line luminosity is estimated
from only a limited number of emission lines.  Since the relative
strengths of the main strong lines are relatively well determined, and
quasars generally have fairly uniform spectra (e.g., Francis et al. 1991;
Zheng et al. 1997), this should not introduce a large source of uncertainty.
More worrisome is our
assumption of a constant cloud covering factor (10\%),
which in reality is likely to have an intrinsic dispersion.  And third,
even if we assume that the host galaxy magnitudes are perfectly measured,
the $M_R -M_{\rm BH}$ relation still has a scatter of $\sim 0.4$ dex
(McLure \& Dunlop 2002). For concreteness, we assume that any individual
value of $\lambda$ may have an uncertainty of $\sim 0.5$ dex.  Although
this is not negligible, fortunately it is not sufficiently large to obscure
gross statistical trends in our sample.  However, we caution against a
literal interpretation of the individual values of $\lambda$ and $\Lambda$
listed in Table 3.

The three critical values of $\lambda$ define four regimes in accretion states,
as shown in Figure~2{\it b}.  The average values of $\Lambda$ for FSRQs and
SSRQs are, respectively, $\langle \log \Lambda \rangle_{\rm FSRQ} =24.0$ and
$\langle \log \Lambda \rangle_{\rm SSRQ}=24.1$.  The corresponding values for
$\lambda$ are $\langle \log \lambda \rangle_{\rm FSRQ}=-2.20$ and
$\langle \log \lambda \rangle_{\rm SSRQ}=-2.05$.  Given the uncertainties just
discussed, the two subsamples appear to be statistically indistinguishable.

Wang et al. (2002) present the distributions of $\Lambda$ and $\lambda$ for
BL Lac objects and conclude that {\it all} BL Lacs have ADAFs. Comparing those
objects with the RLQs presented here, we find that the distributions of
$\Lambda$ overlap significantly, increasing smoothly from BL Lacs to RLQs.
This was noted by Scarpa \& Falomo (1997), who used only \mgii\ as the
luminosity indicator.  Our result considerably strengthens theirs.  A
similar behavior is seen in the distribution of $\lambda$.  As we discuss
later, these trends suggest that an evolutionary sequence, characterized by
different accretion states, may connect BL Lac objects to RLQs.

The normalized mass accretion rates, $\dot{m}=\xi^{-1}\lambda\approx10\lambda$,
span over a relatively large range, from $10^{-2}$ to $10^{0.7}$. Some of the
FSRQs in our sample fall in the range of objects studied by Cavaliere \& D'Elia
(2002), but we suggest that some of FSRQs, like SSRQs, may have hybrid ADAF+SS
disks. With the similarity in $\dot{m}$ between FSRQs and SSRQs, the only
difference between FSRQs and SSRQs is geometrical orientation, which has been
suggested (Urry \& Padovani 1995).

Cavaliere \& D'Elia (2002) suggest that an evolutionary sequence may be defined
for blazars, depending on their dimensionless accretion rates. FSRQs have
$\dot{m}\approx 10^{-1}$, whereas BL Lac objects have $\dot{m}\approx 10^{-3}$,
the two populations separated by a significant gap.
By contrast, our analysis suggests that BL Lacs and RLQs define a
{\it continuous}\ sequence in accretion rate.  Although our conclusion is
based on two heterogeneous samples, we note that it is in general agreement
with the evolutionary scenario proposed by B\"ottcher \& Dermer (2002).

The states of the accretion disk listed in Table 3 can be tested by Fe
K$\alpha$ line observations. Differences in the X-ray properties between
radio-loud and radio-quite quasars have been discussed by Sambruna, Eracleous,
\& Mushotzky (1999) and Reeves \& Turner (2000). The most important findings
in these studies is that the Fe K$\alpha$ line tends to have complex
profiles.  Only a few RLQs have broad Fe K$\alpha$ profiles, most being
undetected (Sambruna et al. 1999; Reeves \& Turner 2000). There is also a
clear trend that more radio-loud objects tend to have lower Fe K$\alpha$
equivalent widths.  By analogy with the interpretation for the relativistically
broadened Fe lines in Seyfert nuclei (e.g., Nandra et al. 1997), RLQs with
strong broad Fe K$\alpha$ lines may be in the SS state, while those lacking
such a feature may be in a pure ADAF or in an ADAF+SS state, according to the
classification suggested in this study.  This interpretation, although
appealing, may not be unique.  Recall that among Seyfert galaxies and
radio-quiet quasars there exists an inverse correlation between Fe line
strength and X-ray luminosity (Iwasawa \& Taniguchi 1993), an effect that
can be attributed to luminosity-dependent changes in the ionization state of
the surface of the disk (Reeves \& Turner 2000).  Alternatively, perhaps
strong relativistic jets physically modify the structure of the innermost
part of the disk, thereby altering or suppressing the iron-line emitting
region (see \S~4.1.3).

We note that values of $\lambda_1$ and $\lambda_2$ are sensitive to
$\alpha$. The most likely value of $\alpha$ is 0.3 (Narayan, Mahadevan
\& Quataert 1998). However if $\alpha=0.1$, it will follow that
$\lambda_1=1.91\times 10^{-4}\xi_{-1}\alpha_{0.1}^2$ and
$\lambda_2=1.0\times 10^{-3}\xi_{-1}\alpha_{0.1}^2$. All the disks listed as
hybrid models of SS and ADAF will be SS disk in Table 3.
In any case, this does not affect the conclusion that most of the objects have
the standard accretion disks.

There is only one possible super-Eddington accretor (PKS 0454$-$22) in the
present sample. This nicely agrees with the result from an independent test
by Wang (2003), in which super-Eddington accretors have been searched in the
present sample and no super-Eddington accretor has been found. He used the
theoretical spectra emergent from slim disks (Wang et al. 1999)
and the empirical reverberation
relation to find a limit relation bewteen the BH mass and H$\beta$ width as
$M_{\rm BH}=(2.9 - 12.6)\times 10^6M_{\odot}
\left(v_{\rm FWHM}/10^3{\rm km~s^{-1}}\right)^{6.67}$ in AGNs with
super-Eddington rates.  With the help of equation (1),
the BH masses can be obtained, he found all the objects in the present sample
are located below this limit.

It is worth noting that, with the exception of 3C~273, our sample lacks very
powerful FSRQs.  This is an observational selection effect.  It is difficult
to measure the host galaxies of the most powerful quasars because of their
very bright nuclei.  We will later return to the issue of whether there exists
super-Eddington RLQs that appear as very powerful FSRQs.

\section{Discussion}

\subsection{Jet Formation from the Disk}

Two processes have been widely advocated for jet formation in AGNs: (1) energy
extraction from the spin of the BH (Blandford \& Znajek 1977, hereafter BZ)
and (2) energy extraction from a disk wind (Blandford \& Payne 1982, hereafter
BP). Several variants of the BZ and BP models have been proposed. The results
of the present study allow us to set some constraints on the theoretical
models. In their investigation of the BZ process in the regime of the SS disk,
Livio, Ogilvie, \& Pringle (1999) conclude that both the BZ and BP power, even
for Kerr BHs, are negligible compared with the radiative output from the
accretion disk itself.  The reason is that geometrically thin disks cannot
produce a strong enough poloidal magnetic field for the energy extraction
to be effective.

\subsubsection{Model of Ghosh \& Abramowicz}

Ghosh \& Abramowicz (1997) assume that the poloidal magnetic field can be
parameterized as $B_p^2\propto \alpha P_{\rm max}$, where $P_{\rm max}$ is the
maximum pressure in the inner region if the accretion rate
$\dot{m}>10^{-3}(\alpha_{-2}m_9)^{-1/8}$, with $\alpha_{-2}=\alpha/10^{-2}$
and $m_9 = M_{\rm BH}/10^9M_{\odot}$.
In such a region, the radiation pressure dominates over the gas pressure and
is independent of the accretion rate. Then, the BZ power is given by

\begin{equation}
L_{\rm BZ}=2\times 10^{45}m_9j^2~~{\rm erg~s^{-1}},
\end{equation}

\noindent
where $j=J_{\rm BH}/(GM_{\rm BH}^2 c^{-1})$ is the specific angular momentum. We
can see that the BZ power is {\it independent} of the accretion rate, but very
sensitive to the spin.  The extended radio power is expected to be proportional
to $L_{\rm BZ}$ (Meier 1999).  For our sample, Figure~3 shows that the extended
radio powers span $\sim$ 5 orders of magnitude, whereas the BH masses are
confined to within roughly a factor of 10.  This means that the BHs should
have spins in the range $10^{-2}<j<1.0$. Such a large range of values for the
spin, however, is not expected in RLQs (Wilson \& Colbert 1995).  The model of
Ghosh \& Abramowicz (1997), therefore, seems to be disfavored by the present
data. Additionally, the results of Lacy et al. (2001) from FIRST Bright Quasar
Survey show that the radio emission depends on the accretion rate
($\propto \dot{m}^{1.0}$), which does not support Equation (10). As pointed
out by Livio et al. (1999), Equation (10) overestimates the BZ power because
the poloidal magnetic field should be $B_p\approx (H/R)B_{\phi}$ rather than
simply $B_p^2/8\pi \approx P_{\rm max}$.

\subsubsection{Model of Meier}

Meier (2001) considers energy extraction from the spin of the BH and from the
disk itself, for both an SS disk and an ADAF. His model uses the middle-region
solution for the SS disk and a self-similar solution for the ADAF; it considers
both Schwarzschild and Kerr BHs. In this model the jet power for a Schwarzschild
BH is only $L_{\rm jet}=10^{41.7}$ \lum\ if $M_{\rm BH}=10^9
M_{\odot}$ and $\dot{m}=0.1$, whereas $L_{\rm jet}=10^{42.7}$ \lum\ for a Kerr
BH with the same mass and accretion rate. The present data clearly do not support
this model because many RLQs have SS disks that support much higher luminosities.
When the accretion disk is in an ADAF state, for example, an accretion rate of
$\dot{m}=10^{-2.5}$ onto a Schwarzschild BH radiates $\sim 10^{42.3}$~\lum,
which may generate a radio power of $10^{24.3}\epsilon_{-2}$W~Hz$^{-1}$ at
1 GHz in the extended lobes, where $\epsilon_{-2}=\epsilon/10^{-2}$ is the
fraction of the total jet power radiated in the radio band (Meier 1999). This
can account for the radio emission from BL Lac objects shown in Figure~3;
it also implies that the BHs of BL Lac objects may not necessarily have large
spins.  The jet power from an ADAF surrounding a Kerr BH is approximately
$10^{44.1}$ \lum, corresponding to a radio power of
$10^{26.1}\epsilon_{-2}$W~Hz$^{-1}$ at 1~GHz. This can account for RLQs in the
ADAF+SS regime.  But how do we explain those RLQs that have much higher
radio powers and that contain pure SS disks?

\subsubsection{An Improved Model}

From the above discussion, it is evident that the power output of the jet,
whether generated through the BZ or BP processes, depends both on the model
adopted for the accretion disk and the parameterization used for the poloidal
magnetic field.  Below we derive an improved formula for the jet power.

Following Meier (2001), the magnetohydrodynamic (MHD) power of the jet is
given by

\begin{equation}
L_{\rm jet}=\frac{1}{32c}B_p^2R_0^4\Omega_0^2,
\end{equation}

\noindent
where $B_p$ is the poloidal magnetic field and $\Omega_0$ is the Keplerian
angular velocity at the radius $R_0$, where the jet develops.  Both the
BZ and the BP process need a strong poloidal magnetic field $B_{p}$, which
is produced by the toroidal magnetic field $B_{\phi}$. Livio et al. (1999)
showed that

\begin{equation}
B_{p}\approx \left(\frac{H}{R}\right)^nB_{\phi},
\end{equation}

\noindent
where $H$ is the height of the disk at radius $R$ and the index $n\approx 1$.
Since $H/R \ll 1$ for an SS disk, it is generally believed that $B_p \ll
B_{\phi}$, and hence that the extracted energy must be small (Livio et al.
1999; Meier 1999, 2001). However we should note that $H/R \approx 0.01$ is
only valid in the middle and outer regions of the disk; in the inner region,
$H/R \approx 1$.  Meier (2001) argues that this region is thermally unstable.
He only uses the middle-region solution in his calculations, and
consequently obtains a reduced power from the BZ process.

The thermal instability arises from insufficient cooling. It is well known that
such an instability in the innermost regions of the disk can be suppressed by
advection, which acts as an efficient cooling process (Abramowicz et al.
1988; Chen \& Taam 1993). Advection mainly converts the heat of the accreting
gas into entropy, rather than radiation.  Consequently, the sound speed
($c_s$) and the scale height $H$ are increased.  It would be of interest to
include the role of advection cooling in the calculation of the BZ power.

We follow Chen's (1995) treatment of advection cooling in determining the
structure of the disk.  The disk is assumed to undergo Keplerian rotation
at an angular velocity $\Omega_{\rm K}$, and we assume conservation of mass,
conservation of angular momentum, and vertical equilibrium. Advection is
included as local cooling, and we neglect the corona.  The energy balance
between the heat production ($Q_{\rm vis}$) and cooling can be written as

\begin{equation}
Q_{\rm vis}=Q_{\rm rad}+Q_{\rm adv},
\end{equation}

\noindent
where $Q_{\rm rad}$ and $Q_{\rm adv}$ represent radiative and advection
cooling, respectively.  The viscous heating rate per unit area is
$Q_{\rm vis}=3\dot{M}\Omega^2fg/4\pi$, where the correction terms $f\approx 1$
and $g\approx 1$ are due to the inner boundary condition and the
pseudo-Newtonian potential (Chen 1995).  Advection cooling can be expressed as
$Q_{\rm adv}= \dot{M}c_s^2\zeta/2\pi R^2$, $\zeta=5.5$ being the advection
factor.  With the diffusion approximation and the assumption of radiation
pressure dominance, we have $Q_{\rm rad}=4acT^4/3\kappa_{\rm es}\Sigma=
2c\Omega_{\rm K}\cs/ \kappa_{\rm es}$, where $\Sigma$ is the surface density
of the disk, $a=7.56\times 10^{-15}$ is the radiation density constant and
$\kappa_{\rm es}$ = 0.34 is the electron-scattering opacity.  With the help of
the equations for mass and angular momentum conservation and vertical
equilibrium, Equation (13) reduces to

\begin{figure*}[t]
\centerline{\includegraphics[angle=-90,width=16.5cm]{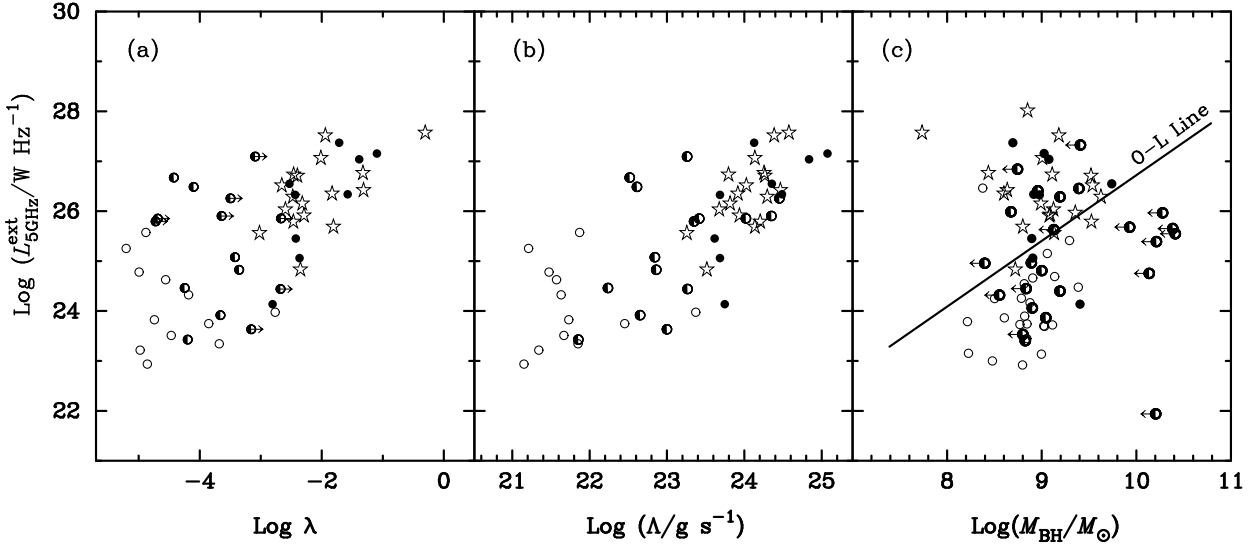}}
\caption{\footnotesize
The distribution of extended radio power at 5~GHz versus the ({\it a})
$\lambda$, ({\it b}) $\Lambda$, and ({\it c}) $M_{\rm BH}$. The filled circles
show the FSRQs, and SSRQs are denoted as open stars. The open and semi-filled
circles represent high frequency-peaked and low frequency-peaked BL Lacs,
respectively; the radio data for these objects are taken from Urry et al.
(2000), and values for $\lambda$, $\Lambda$, and $M_{\rm BH}$ are taken from
Wang et al. (2002). The diagonal line in panel ({\it c}) is the empirical
division between FR~I and FR~II radio galaxies in the diagram of extended
radio power at 1.4~GHz versus optical {\it R}-band absolute magnitude, as
presented by Ledlow \& Owen (1996); see text for details.
}
\label{fig3}
\centerline{}
\vglue -0.6cm
\end{figure*}

\begin{equation}
\frac{\zeta\dot{M}}{2\pi R^2}\cs^2+
\frac{2c\Omega_{\rm K}}{\kappa_{\rm es}}\cs-
\frac{3\dot{M}}{4\pi}\Omega_{\rm K}^2=0.
\end{equation}

\noindent
The first term represents advection cooling, which is only important at
small radii or for high accretion rates, and the second represents radiative
cooling. Neglecting the first term yields the standard solution for an SS
disk, whereas neglecting the second gives the solution for an optically
thick ADAF (or SEA).  Advection cooling helps to suppress thermal instabilities
that develop in the inner regions of the disk.

The solutions of the disk give the sound speed
\begin{equation}
a_{\rm s}=0.09~\dot{m}_{-1}^{-1}r^{0.5}\left\{
              \sqrt{1+33\dot{m}_{-1}^2r^{-2}}-1\right\},
\end{equation}
where $a_{\rm s}=c_{\rm s}/c$ and $\dot{m}_{-1} = \dot{m}/0.1$, the pressure
\begin{equation}
P=1.76\times 10^7 ~
  \alpha^{-1}\dot{m}_{-1}m_9^{-1}a_{\rm s}^{-1}r^{-3}~~{\rm dyne~cm^{-2}},
\end{equation}
and the vertical half-thickness
\begin{equation}
\frac{H}{R}= 
a_{\rm s}r^{1/2},
\end{equation}
where $r=R/R_g$ and $R_g=GM_{\rm BH}/c^2$.
With the help of $B_{\phi}^2/8\pi\approx \alpha P$ and Equation (12), we have
\begin{equation}
B_p=2.02\times 10^4 
    \dot{m}_{-1}^{1/2}m_9^{-1/2}a_{\rm s}^{1/2}r^{-1} ~~~{\rm Gauss}.
\end{equation}

Following Meier (2001), we take the characteristic radius $R_0=7R_g$ and
$R_0=1.5R_g$ for a Schwarzschild and Kerr BH, respectively, and the angular
velocity of the metric in the Boyer-Lindquist frame
$\omega=-g_{\phi t}/g_{\phi \phi}\approx 0.3j\left(GM_{\rm BH}/c^3\right)^{-1}$.
We finally get the jet power from the accretion system.  For a Schwarzschild
BH,
\begin{equation}
L_{\rm jet}=8.60\times 10^{43}~m_9\dot{m}_{-1}~~{\rm erg~s^{-1}},
\end{equation}
and for a Kerr BH,
\begin{equation}
L_{\rm jet}=1.87\times 10^{45}~m_9\dot{m}_{-1}(1+1.1j+0.3j^2)~~{\rm erg~s^{-1}}.
\end{equation}
Considering that the bolometric luminosity from accretion is $1.38\times
10^{45}m_9\dot{m}_{-1}$ \lum, the ratio of the jet power to accretion power
for a Schwarzschild BH is roughly $4\times 10^{-2}$, whereas for a Kerr BH
it is $\sim$1.0.

The output powers given by Equations (19) and (20) are different from those
found by Ghosh \& Abramowicz (1997) and Meier (2001). First, the energy output
based on Equations (19) and (20) depends on the accretion rate, whereas the
model of Ghosh \& Abramowicz (1997) does not.  Second, the output is much
higher than that given by Meier (2001) for the standard accretion disk model.

The MHD power for a Schwarzschild BH is due to the BP process. For a Kerr BH,
the BP process also depends strongly on the spin because the metric itself
drags the MHD fluid, which then contributes to the rotation of the magnetic
field; thus, Equation (20) covers both the BP and BZ processes. The above
levels of MHD power, which is extracted from the BH and the disk in the
standard SS regime, are sufficient to account for the highest radio
luminosities seen in RLQs.

As shown by Xu et al. (1999), the mass outflow rate from the jet is given by
\begin{equation}
\dot{M}_{\rm jet}\approx \left(\frac{H}{R}\right)^{2\delta -3}\dot{M},
\end{equation}
\noindent
where $\delta$ is a constant from 1.7--3.4. For Kerr BHs, the characteristic
radius $r_0=1.5$ leads to $a_{\rm s}\approx 0.65r^{-0.5}$, and hence
$H/R\approx 0.65$ in the jet-forming region.  We then have
$\dot{M}_{\rm jet}\approx (0.19-0.84)\dot{M}$: the mass outflow rate of the
jet is a large fraction of the accretion rate. If the jet-formation region is
an ADAF, the outflow rate will be even higher than this because
$H/R\approx 1$.  We note that such a large outflow rate may have
implications for observations of the broad Fe K$\alpha$ line in radio-loud
AGNs.  The region of the disk that emits the red wing of the broad line
roughly coincides with the region where the relativistic jet is formed.  If
strong winds accompany jets, they can substantially alter the inner structure
of the disk.  This may be at least partly responsible for the observed
diversity of Fe-line strengths and the complexity of their profiles in
RLQs (Reeves \& Turner 2000).

We note that Equation (21) is {\it not}\ self-consistently obtained from
the disk model. Such a high outflow rate may take place in ADIOS models. Here
we stress that the Fe K$\alpha$ line in RLQs may be strongly affected
by the outflow. The case of 3C 120 adds credibility to the picture that
jet formation influences the innermost region of disk: ``dips in X-ray
emission are followed by ejections of bright superluminal knots in the radio
jet'' (Marscher et al. 2002). The detectability of the Fe K$\alpha$ line in
radio-loud objects may also be influenced by contamination of 
the X-ray continuum by non-thermal (beamed) emission.

\subsection{Empirical Evidence for a Disk-Jet Connection}

The correlation between the extended radio and line emission has been explored
by a number of authors (e.g., Baum \& Heckman 1989a, 1989b; Rawlings \&
Saunders 1991; Rawlings 1992; Falcke, Malkan, \& Biermann 1995; Xu et al.
1999; Cao \& Jiang 1999; Willott et al. 1999). The present sample allows
us not only to further test this correlation, but presents us with an
opportunity to explain it in the context of our improved model.
Figures 3{\it a}\ and 3{\it b}\ show that the extended radio emission is
evidently quite closely linked with $\lambda$ and $\Lambda$. The linear
regression for the RLQs alone is
\begin{equation}
\log L_{\rm 5GHz}^{\rm ext}=(28.05\pm 0.78)+(0.95\pm 0.38)\log \lambda,
\end{equation}
\noindent
with a Pearson's correlation coefficient $r=0.56$, significant at the level of
$p = 3.8\times 10^{-3}$ according to Student's {\it t}\ test.  Similarly,
\begin{equation}
\log L_{\rm 5GHz}^{\rm ext}=(-8.65\pm 11.03)+(1.45\pm 0.46)\log \Lambda,
\end{equation}
\noindent
with $r=0.71$ and $p=6.2\times 10^{-5}$.

For comparison, we include the BL Lac objects studied by Wang et al. (2002)
that have extended radio luminosities given by Urry et al. (2000); there are
12 high frequency-peaked (HBL) and 15 low frequency-peaked (LBL) objects.  The
BL Lac objects smoothly extend the sequence defined by the RLQs toward lower
values of $L_{\rm 5GHz}$, $\lambda$, and $\Lambda$.  The correlations
are now enhanced, becoming
\begin{equation}
\log L_{\rm 5GHz}^{\rm ext}=(27.57\pm 0.29)+(0.68\pm 0.10)\log \lambda,
\end{equation}
\noindent
with $r=0.65$ and $p=2.6\times 10^{-10}$, and
\begin{equation}
\log L_{\rm 5GHz}^{\rm ext}=(4.18\pm 2.70)+(0.92\pm 0.11)\log \Lambda,
\end{equation}
\noindent
with $r=0.76$ and $p=2.1\times 10^{-15}$.  In the above correlation analysis
of $L_{\rm 5GHz}^{\rm ext}$ versus $\lambda$, we excluded the upper limits
for the BL Lac objects. A survival analysis including these upper limits
shows $\log L_{\rm 5GHz}^{\rm ext}=(27.60\pm 0.30)+(0.64\pm 0.11)\log \lambda$
with Spearman's $\rho=0.667$. These correlations are striking in
that they provide fairly direct evidence that the extended jet emission is
somehow closely coupled to the central engine via the accretion disk.  The
most direct interpretation is that the extended radio emission is fed by
relativistic jets that are tightly linked to the accretion disk (Rees 1984).

Following Meier (2001), we assume that 1\% of the MHD energy will be
radiated as radio emission with a spectral index of $-1$.  Then, for a Kerr BH
and using the relation $\Lambda=\eta\xi\dot{M}$,
$L_{\rm 5GHz}^{\rm ext}=3.74\times 10^{26}\epsilon_{-2}m_9\dot{m}_{-1}$
W~Hz$^{-1}$, or
\begin{eqnarray}
L_{\rm 5GHz}^{\rm ext}&=&
3.74\times 10^{27} \epsilon_{-2}m_9\lambda~~ {\rm W~Hz^{-1}}\nonumber \\
~&=&2.69\times 10^2\epsilon_{-2}\Lambda~~{\rm W~Hz^{-1}}.
\end{eqnarray}
These expressions are in rough agreement with the observed correlations.

\subsection{A Power Limit for Radio-Loud Quasars?}

In our sample, only PKS~0454$-$22 is a super-Eddington accretor 
($E_R = 10^{0.7}$). The most powerful RLQs, which have powers in excess 
of $10^{47}$
\lum, potentially may be super-Eddington accretors (Ghisellini \& Celotti
2001).  In this section we consider the upper luminosity limit of RLQs as a
threshold set by super-Eddington accretion.

For a BH accreting well above the Eddington rate, much of the dissipated
energy will be advected into BH because most of the photons are trapped in
the flow by Thomson scattering (Wang \& Zhou 1999; Ohsuga et al. 2002).
Photon trapping efficiently controls the radiated luminosity;
the radiated luminosity from the disk depends very weakly on the accretion
rate, $L_{\rm disk}\propto \log \dot{M}$, tending to be saturated.
For a self-similar disk (Wang \& Zhou 1999), Equation (15) reduces to
\begin{equation}
a_{\rm s}\approx 0.65 r^{-0.5},
\end{equation}
and the MHD power is given by
\begin{equation}
L_{\rm jet}=8.05\times 10^{46}~\dot{m}_{1}m_9~~{\rm erg~s^{-1}}
\end{equation}
for a Schwarzschild BH and
\begin{equation}
L_{\rm jet}=3.76\times 10^{47}~\dot{m}_{1}m_9~~{\rm erg~s^{-1}}
\end{equation}
for a Kerr BH, where $\dot{m}_1=\dot{m}/10$. The power is still proportional
to the accretion rate, although the radiation from disk itself is saturated.
In the super-Eddington regime, the difference in radio power between a Kerr 
and Schwarzschild BH is smaller.

From the total rotational energy due to the spin,
$E_{\rm spin}
          \approx 1.6\times 10^{62} m_9j^2~~{\rm erg}$, it follows that the
spin-down timescale due to the BZ process is

\begin{equation}
\tau_{\rm spin-down}=1.9\times 10^{7}~\dot{m}_1^{-1}~~{\rm yr}.
\end{equation}

\noindent
This is much shorter than a typical merger timescale of $10^9$ yr (Wilson \&
Colbert 1995), if the spin angular momentum comes from galaxy collisions.
If the powerful jet is fed by a Schwarzschild BH via super-Eddington
accretion, the $e$-folding timescale of the BH growth will be
\begin{equation}
\tau_{_{\rm SEA}}=M_{\rm BH}/\dot{M}=4.0\times 10^7\dot{m}_1^{-1}~{\rm yr}.
\end{equation}
The lifetime of the powerful jet from an SEA is rather short unless there is
continuous super-Eddington fueling. Thus, super-Eddington RLQs ought to be
very rare.

\subsection{AGN Unification}

A useful diagnostic for radio galaxies, as first shown by Owen \& Ledlow
(1994; see also Ledlow \& Owen 1996 and Owen, Ledlow, \& Keel 1996), is the
plot of extended radio luminosity versus optical absolute magnitude.  A
diagonal line well separates the two radio populations: at a given
optical luminosity (which in the case of radio galaxies is dominated by
the host galaxy), FR~IIs have much higher radio powers than FR~Is (Fanaroff
\& Riley 1974).  The physical interpretation of the Owen-Ledlow line has been
much debated.  Bicknell (1995) and Gopal-Krishna \& Witta (2001) suggest that
it may arise naturally from the interaction of the jet with the interstellar
or intracluster medium. Or perhaps the two classes reflect intrinsic
differences in their central engines, such as BH spin (Baum, Zirbel, \& O'Dea
1995; Meier 1999) or state of the accretion disk. Ghisellini \& Celotti (2001)
recently suggested that the dividing line corresponds to a transition in the
accretion mode, from an SS disk (FR~II) to an optically thin ADAF (FR~I).

Urry et al. (2000) also examined BL Lac objects and RLQs in the
$M_R - L^{\rm ext}_{\rm radio}$ diagram, but the data for their RLQs are
mostly not from {\it HST}\ observations. Figure 3{\it c}\ gives a new
representation of the Owen-Ledlow diagram for RLQs and BL Lacs, where we 
have substituted the optical luminosity on
the abscissa with BH mass $M_{\rm BH}$. We have plotted our sample of RLQs
along with the sample of BL Lac objects studied by Urry et al. (2000). The
diagonal line is the Owen \& Ledlow (1994) line for radio galaxies, adapted
from Ledlow \& Owen (1996) after converting to our distance scale, the radio
luminosities from 1.4~GHz to 5~GHz assuming a spectral index of $-1$, and
$M_R$ to $M_{\rm BH}$ using Equation (1).

Several features are noteworthy in this diagram.  First, the Owen-Ledlow line
that divides FR~I and FR~II radio galaxies seems to cleanly bisect the
population of beamed radio-loud objects.  Nearly all HBLs fall below the line,
and all but two FSRQ lies above the line.   Second, the population of LBLs
straddle the divide, occupying the region between HBLs and FSRQs. Third,
as with the FSRQs, essentially all SSRQs line above the line. And
fourth, at least some BL Lac objects, mostly of the LBL variety, appear to be
more closely associated with FR~II radio galaxies; this is in agreement with
the suggestion of Kollgaard et al. (1996) that some radio-selected BL Lac
objects may be beamed FR II sources.






The availability of accretion rates enables us to consider radio-loud AGNs in
a more unified, physical context.  Wang et al. (2002) show that most
BL Lac objects have accretion rates below the critical value
$\dot m \approx \alpha^2$ that demarcates the transition from a standard
SS disk to an optically thin ADAF.  Moreover, they find that HBLs have
characteristically lower accretion rates than LBLs.  This study demonstrates
that nearly all RLQs have $\dot m > \alpha^2$.  It is important to note
that, at least for the sample considered here, most of the sources occupy a
relatively narrow range of BH masses, $\sim 10^{8.5}$ \solmass\ with a spread
of $\sim$0.5 dex.  The various subpopulations of radio-loud AGNs can therefore
be cast into a physical, perhaps evolutionary, sequence described by a single
variable --- accretion rate.  For the beamed population, $\dot m$ decreases
along the sequence FSRQ $\rightarrow$ LBL $\rightarrow$ HBL.  When viewed at
large angles to our line of sight, the sequence becomes
FR~II $\rightarrow$ (FR~II or FR~I) $\rightarrow$ FR~I.  Since the structure
of the accretion flow is governed by $\dot m$, the variation in $\dot m$ along
the sequence directly translates into a variation in the dominant mode of
accretion.  The most powerful RLQs and FR~II radio galaxies may contain
super-Eddington disks or optically thick ADAFs.  The majority of RLQs and
FR~IIs have SS disks. The low-power HBLs and some LBLs have pure optically thin
ADAFs.  Objects that occupy the fuzzy boundary near the Owen-Ledlow line,
which comprise some LBLs and a minority of FR~IIs, may have a hydrid structure
consisting of an SS disk plus an optically thin ADAF.

The basic elements of the above evolutionary scenario, based on complementary
but different lines of evidence, have already been advocated by a number of
authors (e.g., Ghisellini et al. 1998; Ghisellini \& Celotti 2001, 2002;
Ghisellini, Celotti, Costamante 2002; B\"ottcher \& Dermer 2002; Cavaliere \&
D'Elia 2002).
B\"ottcher \& Dermer (2002) and Cavaliere \& D'Elia (2002) argue
that RLQs evolve into BL Lacs rather than BL Lacs into RLQs (B\"ottcher \&
Dermer 2002).
If the evolutionary direction is from RLQs to BL Lacs, the BH masses of
RLQS should be  systematically smaller than those of BL Lacs. However, the
significant overlap in the BH mass distributions of the two groups
(Fig. 3{\it c}; see also Falomo 2003) does not seem to support this single
evolutionary direction.
Given the current uncertainties in BH mass determinations for AGNs, it is
probably premature to draw strong conclusions.  Here, we merely wish to note
that {\it both}\ evolutionary paths can, and probably do, occur.  The
detection of quasar remnants in nearby galaxies in the form of inactive
massive BHs conclusively demonstrates that the cosmic evolution of AGN
activity must be episodic (Richstone et al. 1998).  The duty cycle for
accretion is short, such that any individual massive BH is likely to have
been activated (and deactivated) many times since it was formed.

\section{Conclusions}

We attempt to constrain the accretion rates and the mode of accretion in
RLQs using BH masses estimated from {\it HST}\ imaging of their host galaxies
and accretion luminosities derived from their emission-line spectra.  The
quasars have large BH masses, $M_{\rm BH} \approx 10^8-10^9$ \solmass, and
substantial accretion rates, $\dot M \approx 0.01-1$ times the Eddington rate.
Their accretion rates suggest that most of the objects possess standard
optically thick, geometrically thin accretion disks, perhaps in combination
with an optically thin ADAF in some cases.  Our sample has a deficit of
objects with super-Eddington disks, although this is likely due to  a
selection effect in the sample.

Our data can also be used to test current ideas on jet formation. We
present an improved model for a standard optically thick, geometrically
thin disk that incorporates the effects of advection cooling.  Our model can
explain the level of radio emission observed in RLQs.  We discuss some
consequences of our model for the interpretation of Fe K$\alpha$ lines observed
in the X-rays.

Finally, we combine the present sample of RLQs with a recently analyzed sample
of BL Lac objects to reevaluate the unification picture for radio-loud AGNs.
The strong correlation found between the luminosity of the extended radio
emission and the accretion rate supports the notion that radio jets are
directly coupled to the accretion disk.  In agreement with prevailing ideas,
our results support the proposition that FSRQs and BL Lacs are the beamed
counterparts of FR~II and FR~I
radio galaxies, respectively.  However, we argue that some fraction of the low
frequency-peaked BL Lac sources in fact may be more closely associated with
FR~II radio galaxies.  The various subclasses of radio-loud AGNs --- ranging
from high-luminosity sources (FSRQs and FR~IIs) on the one end to
low-luminosity sources (BL Lacs and FR~Is) on the other --- can be viewed as
a continuous sequence of varying accretion rate.

\acknowledgements{The authors are grateful to the referee for helpful
comments and suggestions.  J.~M.~W. thanks D.~L. Meier for useful discussions. 
He is grateful to the
support from Alexander von Humboldt Foundation, the Hundred Talent Program of
Chinese Academy of Sciences, and the Special Funds for Major State Basic
Research Projects and NSFC. The research of L.~C.~H. is supported by the
Carnegie Institution of Washington and by NASA grants from the Space Telescope
Science Institute (operated by AURA, Inc., under NASA contract NAS5-26555).
}

\end{document}